# Tuning of the Thermoelectric Figure of Merit of $CH_3NH_3MI_3$ (M=Pb,Sn) Photovoltaic Perovskites

*Xavier Mettan*, Riccardo Pisoni, Péter Matus, Andrea Pisoni, Jaćim Jaćimović, Bálint Náfrádi, Massimo Spina, Davor Pavuna, László Forró and Endre Horváth*

Laboratory of Physics of Complex Matter, EPFL, CH-1015 Lausanne, Switzerland.

*xavier.mettan@epfl.ch

## ABSTRACT

The hybrid halide perovskites, the very performant compounds in photovoltaic applications, possess large Seebeck coefficient and low thermal conductivity making them potentially interesting high figure of merit (*ZT*) materials. For this purpose one needs to tune the electrical conductivity of these semiconductors to higher values. We have studied the $CH_3NH_3MI_3$ (M=Pb,Sn) samples in pristine form showing very low *ZT* values for both materials; however, photoinduced doping (in M=Pb) and chemical doping (in M=Sn) indicate that, by further doping optimization, *ZT* can be enhanced toward unity and reach the performance level of the presently most efficient thermoelectric materials.

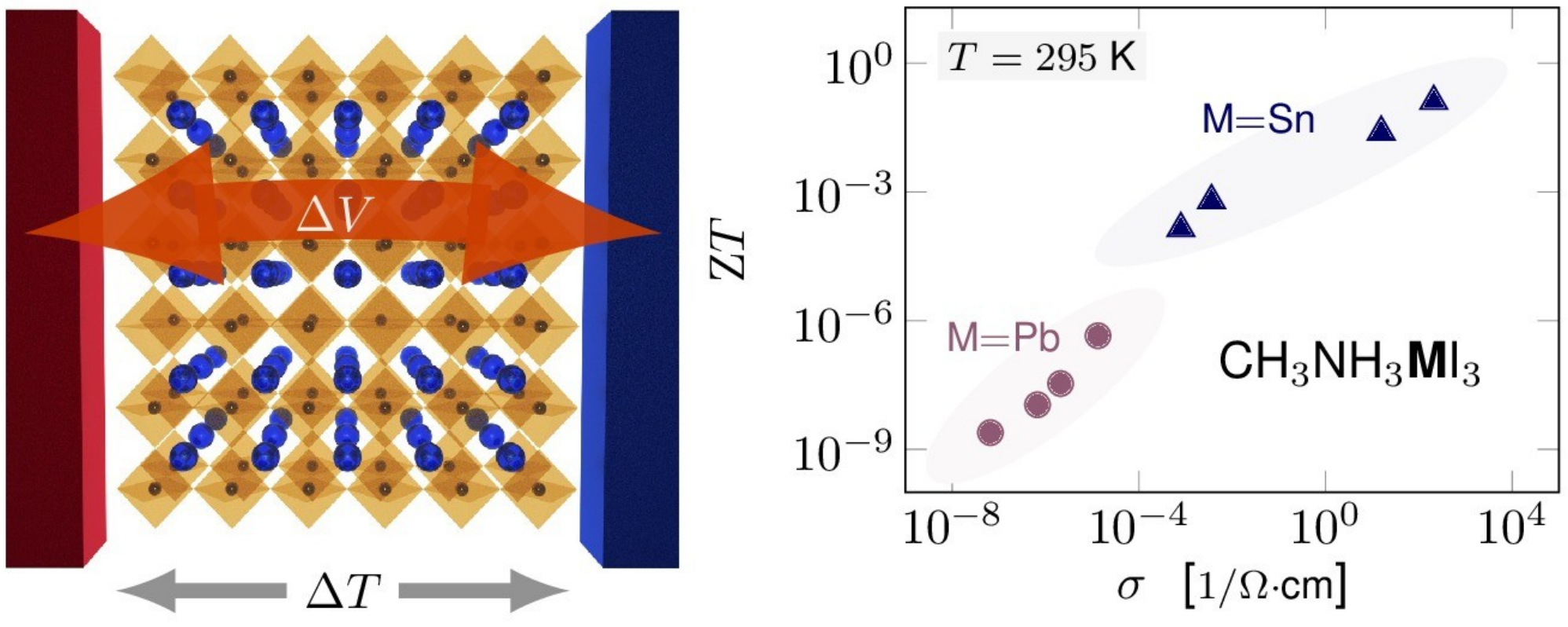

## I. INTRODUCTION

Thermoelectric materials have great potential for future device applications in energy conversion because they transform a temperature gradient, imposed across their volume, into electricity.[1,2] They are especially foreseen to harvest electrical current from waste-heat production.[3] Solar thermoelectric generation is an emerging technology combining concentrated solar power and thermoelectric effect.[4] Here the solar light is collected to generate the temperature gradient across the active thermoelectric material. The success of this technique strongly relies on the efficiency of the thermoelectric materials involved. The efficiency of thermoelectric devices is determined by a dimensionless parameter called figure of merit ($ZT$), which corresponds to the energy conversion efficiency of a Carnot cycle.[1] $ZT$ is defined as $(S^2\sigma/\kappa)T$, where $T$ is the absolute temperature, $S$ is the Seebeck coefficient (or thermoelectric power), $\sigma$ is the electrical conductivity and $\kappa$ the thermal conductivity. The best known thermoelectric materials have a $ZT$ in the range of 2.[5] Nevertheless, synthesizing materials with a higher $ZT$ would be highly beneficial. Therefore, there is a constant search for engineering thermoelectric materials with a high $ZT$, in addition to low production costs. Lately, hybrid halide perovskites $CH_3NH_3PbI_3$ (hereafter $MAPbI_3$) and $CH_3NH_3SnI_3$ (hereafter $MASnI_3$) have been considered for thermoelectric applications, since they possess an extremely large Seebeck coefficient and an unusually low thermal conductivity.[6] Theoretical modeling of He and Galli[7] have foreseen that with suitable doping the $ZT$ value of these two compounds could raise into the range required for thermoelectric applications. In this work, we systematically investigate the transport properties and the figure of merit of potentially low-cost and solution-processable $MAPbI_3$ and $MASnI_3$ samples, studying their evolution with photo- and impurity doping.

## II. EXPERIMENTAL SECTION

**$MAPbI_3$ and $MASnI_3$ Synthesis**. Single crystals of $MAPbI_3$ and $MASnI_3$ were prepared by solution growth. $MAPbI_3$ single crystals were prepared by precipitation from a concentrated aqueous solution of hydriodic acid (57 w% in $H_2O$, 99.99 % Sigma-Aldrich) containing lead (II) acetate trihydrate (99.999 %, Acros Organics) and a respective amount of $CH_3NH_2$ solution (40 w% in $H_2O$, Sigma-Aldrich). A constant 55-42 °C temperature gradient was applied to induce saturation of the solute in the low temperature region of the solution. After 24h, submillimeter sized flake-like nuclei were floating on the surface of the solution (see supporting information of ref 8). Large cube-like and rhombohedral $MAPbI_3$ crystals with 3-5 mm silver-gray facets were obtained after 7 days (left inset of Figure 1). $MASnI_3$ crystals were prepared by precipitation from a concentrated aqueous solution of hydriodic acid (57 w% in $H_2O$, 99.99 % Sigma-Aldrich) containing tin (IV) iodide and a respective amount of $CH_3NH_2$ solution (40 wt% in $H_2O$, Sigma-Aldrich). Tin (IV) iodide was previously prepared by the reaction of tin (II) chloride (99.99 % Sigma-Aldrich) and a respective amount of hydriodic acid (57 wt% in $H_2O$, 99.99 % Sigma-Aldrich). The bright orange solid was subsequently filtered and dried at 80°C. Usually $MASnI_3$ crystals are of smaller size, and very often heavily twinned (right inset of Figure 1). To prevent any unwanted reaction of the samples with moisture and air, the as- synthesized crystals were immediately transferred and kept in a desiccator prior to the measurements. Because we observed a possible dependency of $S$, $\rho$, and $\kappa(T)$ on the environmental conditions, every measurement was performed under high-vacuum ($p \leq 5\cdot10^{-5}$ mbar). In this way, possible undesirable ionic conduction induced by water, as well as thermal dissipation in the surrounding gas were avoided. The $MASnI_3$ crystallites were compressed into pellets to allow a more reliable measurement of the transport coefficients, while the $MAPbI_3$ crystals were directly connected for transport measurements.

**Electrical and Thermal Transports Characterization.** A conventional four-point technique was employed to measure the resistivity. Gold wires were glued on the sample with carbon-containing paste to minimize the Schottky barrier at the interface. For the measurement of the Seebeck coefficient, a temperature gradient was imposed by a resistive heater attached to one extremity of the sample, driven by an alternating current of frequency $f$ = 60 mHz. The temperature gradient across the sample was measured by a Chromel-Constantan (Type E) differential thermocouple, using a lock-in amplifier at 2*f*. The temperature gradient was carefully set to be smaller than 1 K. The thermoelectric voltage was amplified by a Keithley 6517 electrometer and determined using a lock-in amplifier at 2*f*. This measurement was performed on polycrystalline samples, as well, without noticeable difference with the values obtained on single crystals. The thermal conductivity of $MASnI_3$ was measured in a configuration previously described elsewhere,[6,9] whereas for $MAPbI_3$, data are recovered from a previous publication of our group.[6] κ was measured by a steady-state method, using calibrated stainless steel as reference sample. The temperature gradient across the sample was kept around 1 K, and it was measured with type E differential thermocouples connected to the sample and the reference with Stycast 2859 Ft thermal conducting epoxy. The electrical resistivity of the same $MAPbI_3$ crystal was measured in the dark and under white-light illumination of 80, 165 and 220 $mW/cm^2$ (In Figures 1 and 2: red, orange, and yellow curves, respectively). Measurements of the Seebeck coefficient were performed on the same sample in dark and under white-light emanating from a light-emitting diode (LED). $MASnI_3$ crystals from different syntheses were employed for measurements of $\rho$. The Seebeck coefficient of $MASnI_3$ was measured for the most resistive sample. No light intensity dependency was observed in the $\rho$ and $S$ values for $MASnI_3$; therefore all of the measurements were conducted in dark. The thermal conductivity of $MASnI_3$ was measured on the same sample as the one studied for Seebeck coefficient. We investigated a temperature range from 4 to 300 K, for every measurement of $S$, $\rho$, and $\kappa$; however, because of the high resistance of some of the samples, we were often limited to narrower temperature ranges (e.g., for $S$ of $MAPbI_3$ and $MASnI_3$ and for $\rho$ of $MAPbI_3$ in dark and of $MASnI_3$ at low doping levels).

## III. RESULTS AND DISCUSSION

The temperature-dependent transport coefficients $S$, $\rho$, $\kappa$, and the resulting $ZT$ are reported for both $MAPbI_3$ and $MASnI_3$ in Figure 1. The resistivity and thermoelectric power of $MASnI_3$ have already been explored recently,[10-12] whereas the resistivity and thermal conductivity of $MAPbI_3$ in dark have been discussed in ref 6. The resistivity of $MAPbI_3$ (Figure 1b) in the dark is smoothly decreasing with increasing temperatures ($d\rho/dT < 0$), reaching $\rho(295\text{ K}) = 15$ MΩ cm. Upon white-light illumination the resistivity globally decreases up to more than two orders of magnitude as an increasing number of photoexcited carriers become available for conduction. When the intensity of white light increases, the absolute value of ρ decreases. The kink at 150 K accounts for the structural phase transition from a tetragonal to an orthorhombic symmetry of the lattice. Furthermore, the effect of the phase transition can also be seen in the thermoelectric power (Figure 1a) and thermal conductivity temperature dependencies (Figure 1c). The observed small shift in the temperature of this transition might be due to some heating effect of the lamp. At $T < 150$ K, $\rho$ of the illuminated sample decreases with lowering temperature in a metal-like behavior. This unexpected behavior will be addressed in more detail in a forthcoming work (unpublished results).

The temperature dependency of the Seebeck coefficient of $MAPbI_3$ follows a nontrivial behavior with peaks appearing at $T \simeq 270$ , $\simeq 215$ and $\simeq 190$ K in the dark and at $\simeq 280$ and $\simeq 210$ K under white light illumination. The absolute value of the thermoelectric power of $MAPbI_3$ was found to be slightly sample-dependent; however, the overall temperature behavior is similar from one sample to another. The positive Seebeck coefficient argues for holes as the dominant type of charge carriers, but its non-monotonic temperature dependency indicates that electrons, whose temperature-dependent mobility differs from the one of holes, also participate in the charge transport. Independent variations of the contributions of holes (or electrons) weighted by their respective mobilities to the Seebeck coefficient could be the reason for some of the peaks. The decrease in S at room temperature (295 K) upon illumination, from $S_{dark} = 0.82$ mV/K to $S_{light} = 0.54$ mV/K, is consistent with the resistivity data because electrons are photoinduced in the conduction band, decreasing the absolute value of $S$.

The temperature dependent resistivity of $MASnI_3$ samples (Figure 1f) varies from a thermally activated behavior (with activation energy $E_a$ equal to 0.25 eV) to a metallic one for the different samples investigated. It has been previously reported that $MASnI_3$ is a doped semiconductor.[12] The presence of an unknown but significant amount of impurity levels within the $MASnI_3$ band gap is at the origin of this varying $\rho(T)$. The material has been intentionally doped by Takahashi *et al.,*[10] thus allowing to obtain a metallic behavior. Two of their $\rho(T)$ curves are reprinted there (Figure 1f). The observed tunability of the electrical conductivity in $MASnI_3$ is therefore expected to be like in other extrinsic semiconductors.[13] Thus, by doping $MASnI_3$, it is possible to obtain larger carrier densities and optimize the thermoelectric properties.

The negative sign of $S$ of $MASnI_3$ (Figure 1e) indicates that the majority carriers are electrons, in agreement with a previous study;[14] however, the negative sign is in contradiction with some previous results.[10] This discrepancy might be due to the different nature of the impurity levels present in the band gap. A predominant number of donor over acceptor impurity levels explains the negative sign observed in our sample. The absolute value of $S$(295 K) = -0.72 mV/K is rather high, one order of magnitude larger than the value reported in ref 14, because the samples under consideration are semiconducting. Upon doping, S decreases (not shown), and for the metallic samples, it has a linear temperature dependency.[10] Its value stays in the 100 μV/K range at 295 K.

The thermal conductivity of both compounds is very low, presumably because of the rattling motion of the organic cation, but for $MASnI_3$, $\kappa = 0.09\pm0.01$ W/m K is even lower than the one observed in $MAPbI_3$ ($\kappa = 0.5\pm0.1$ W/m K)[6] (Figure 1g). This could be partially due to the polycrystalline nature of the $MASnI_3$ sample studied; however, the overall temperature dependency of $\kappa$ in $MASnI_3$ is very different from the one in $MAPbI_3$. The broad maximum around 220 K suggests that there are some small structural motives that limit the phonon mean free path already at high temperature, in contrast with $MAPbI_3$ where it happens at 30 K. The overall temperature variation of $\kappa$ in $MASnI_3$ seems to have a glassy behavior, which might stem from an effective scattering source spread over a larger temperature range. The ultralow thermal conductivity value of $MASnI_3$ is the key ingredient for high *ZT*. It is reasonable to assume the same value of $\kappa$ for our $MASnI_3$ samples with different doping levels because a previous study showed that the Wiedemann-Franz electronic contribution to the thermal conductivity is negligible within this class of materials;[6] however, for the more conductive samples of Takahashi *et al.*,[10] an additional contribution of the charge carriers to the thermal conductivity ($\kappa_{el}$) cannot be neglected and has to be added.

The temperature dependency of *ZT* for $MAPbI_3$ (Figure 1d) and $MASnI_3$ (Figure 1h), calculated from the above-reported transport coefficients, increases globally with temperature. For $MAPbI_3$ in dark it reaches a maximum value of $10^{-9}$ only, just below room temperature. Photoinduced doping initiates an elevation of *ZT* because the electrical conductivity strongly increases, while $S$ is maintained to a high value, yet *ZT* keeps very low, indicating that this form of doping is not sufficient for thermoelectric applications. Indeed, Hall-effect measurements yield a carrier density $n \simeq 10^{14}$ cm$^{-3}$ and a carrier mobility $\mu \simeq 13$ cm$^2$ V$^{-1}$ s$^{-1}$ for the strongest illumination condition, whereas theoretical calculations of He and Galli[7] predict that $n \simeq 10^{18}$-$10^{19}$ cm$^{-3}$ is required to bring *ZT* close to unity. In the case of $MASnI_3$, the most resistive sample has the magnitude well above the one of $MAPbI_3$. Intentional (as reported in ref 10) or unintentional (as in our case) doping gives much higher conductivities, so that *ZT* at room temperature reaches the value of 0.13. This tendency is very promising, and it should be explored systematically in the future.

The evolution of *ZT* at room temperature is summarized in Figure 2 as a function of the electrical conductivity. It emphasizes that σ is the quantity which tunes *ZT* independently if its increase is due to the enhancement of the carrier density or of the charge mobility. We suppose that carrier mobility does not change substantially with doping, so this plot gives a good approximation for the variation of *ZT* with *n*, as well. As it can be seen in Figure 2, σ plays a dominant role in the modulation of *ZT*. Indeed, the difference in resistivity, two orders of magnitude between the nonilluminated and the illuminated $MASnI_3$ sample, does not result in a dramatic decrease of the thermoelectric power under the same condition.

For an efficient thermoelectric conversion, and to compete with the best available materials, *ZT* = 3 is required. We speculate that increasing *S*, for example by introducing some polaronic effects[15] can be a rather complicated task and $\kappa$ is already very low. Instead, doping $CH_3NH_3SnI_3$ will be the route to further increase *ZT*. According to the trend shown in Figure 2, one can extrapolate the room temperature resistivity necessary to satisfy this request: $\rho = 1\cdot10^{-3}$ to $3\cdot10^{-5}$ Ω cm. This value is not more than two orders of magnitude lower than that observed in intentionally doped $MASnI_3$ crystals by Takahashi *et al.*[10,12] In their work, they studied the effect of the dopant uptake during the crystal growth, under equilibrium conditions, from $Sn^{II}$ and $Sn^{IV}$ precursors. To quantitatively evaluate the carrier concentrations, they performed Hall effect measurements. They concluded that the dopant uptake was rather low, that is the carrier concentration between the as-grown and the artificially doped $MASnI_3$ samples was increased by one order of magnitude only. However, this "$Sn^{II}$ and $Sn^{IV}$ precursor"- doping can be instead regarded as self-doping. Hence this is not likely an effective way to shift the thermodynamic equilibrium between the $Sn^{II}$ and $Sn^{IV}$ ions in the resulted $MASnI_3$ crystal. Therefore, we suggest to shift and stabilize the $Sn^{II}$ and $Sn^{IV}$ redox equilibrium in the Sn-perovskite by doping with foreign cations having a stable oxidation state. For instance, the room-temperature conductivity of the single crystals of metallic slabs of the radical cation of ethylenedithio-1,2-diiodo-tetrathiafulvalene, EDT-TTF-I2, and polymeric lead iodide covalent anionic layers ($\beta$-(EDT-TTF-$I_2)_2^{\bullet+}[(Pb_{5/6}\square_{1/6}I_2)^{1/3-}]_3$) has increased with two orders of magnitude upon doping with silver.[16] Another path to achieve a higher electrical conductivity in the tin halide perovskite could be a postgrowth reduction of pristine or foreign cation-doped $MASnI_3$.

## IV. CONCLUSIONS

In the last couple of years hybrid halide perovskites have been in the focus of research for photovoltaic systems. $CH_3NH_3PbI_3$ has turned out to be a highly efficient material in converting solar light (photons) into electricity, even in fairly simple device configurations. Recently, interest has been drawn to the fact that the same family of compounds might demonstrate good ability to convert heat into electricity. Thermoelectric devices, if their efficiencies stand high enough, could harvest waste-heat and turn it into electricity or even be utilized in power plants. Thus, they would help to reduce the global consumption of nonrenewable energy. In our study we measured the thermoelectric properties of $CH_3NH_3PbI_3$ and $CH_3NH_3SnI_3$. We show that *ZT* of the latter compound at room temperature can be enhanced by more than three orders of magnitude by chemical doping. Further engineering of this material could allow us to reach or even outclass the performance level of the presently most efficient thermoelectric devices. If, by a suitable doping, one could enhance the conductivity, *ZT* would increase toward unity and create a very performant material for such applications. In addition, $CH_3NH_3SnI_3$ was synthesized from a solution, potentially allowing for cheap and massive production, while it is considered as nontoxic. Biological assays for both compounds are still missing, though.

## ACKNOWLEDGMENT

We thank Eric Bonvin for useful discussion. This work was supported by the Swiss National Science Foundation and partially by the CCEM funded HITTEC project.
The authors declare no competing financial interest.

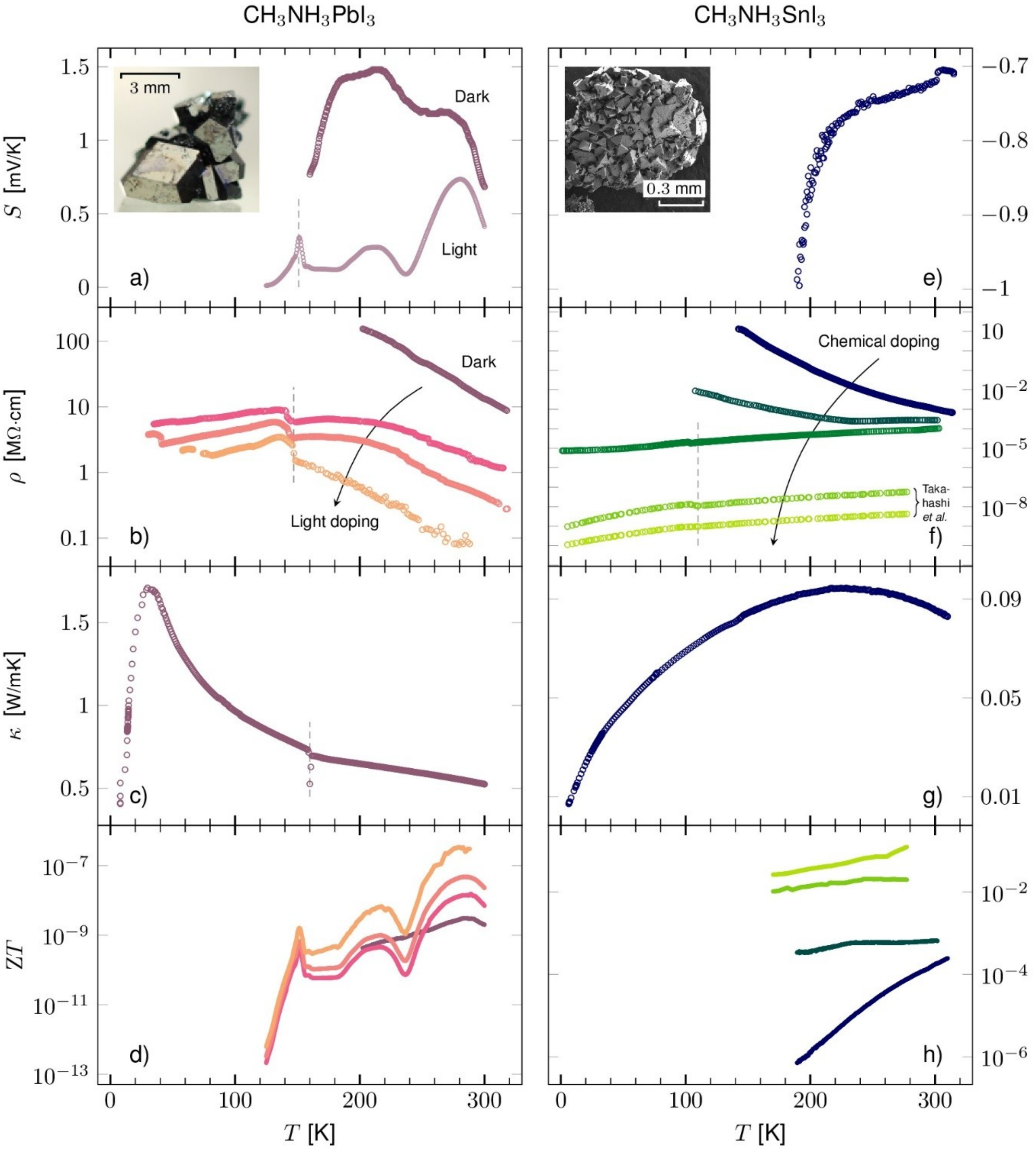

Figure 1 From top to bottom: thermoelectric power, $S$, electrical resistivity, ρ, thermal conductivity, κ, and figure of merit, $ZT$ of $CH_3NH_3PbI_3$ (left panel) and $CH_3NH_3SnI_3$ (right panel) for light (photoelectron) and impurity doping. The three curves of $MAPbI_3$, red, orange, and yellow correspond to light intensities of 80, 165, and 220 mW/cm$^2$. $\kappa$ of $MAPbI_3$ (c) was reproduced from ref 6, and the two lowermost $\rho(T)$ of $MASnI_3$ (f) were reproduced from ref 10. The vertical dashed lines denote a structural phase transition observed in both $MAPbI_3$ and $MASnI_3$. Left inset: Optical image of a $MAPbI_3$ crystal. Right inset: Scanning electron microscope image of a $MASnI_3$ crystallite.

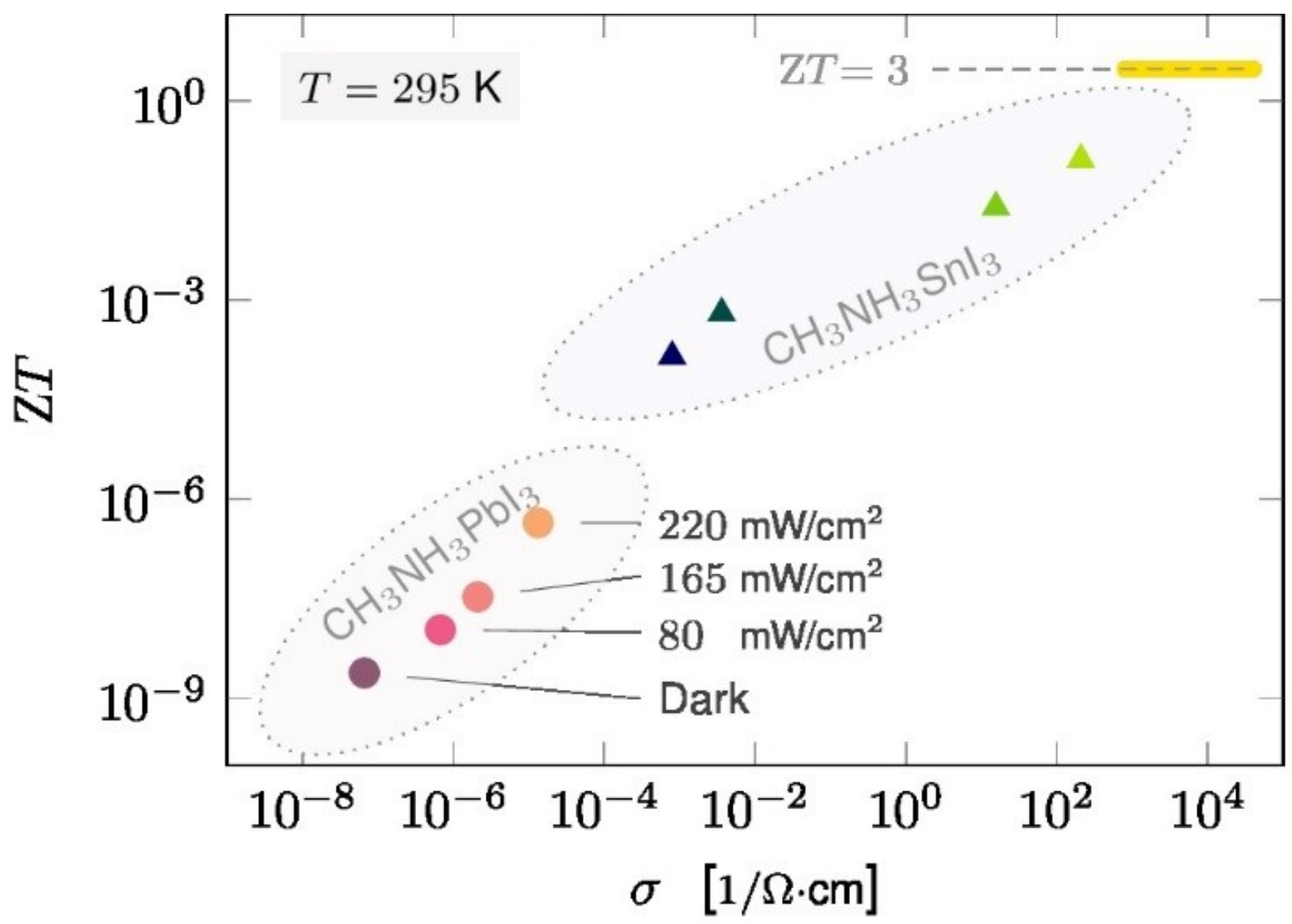

Figure 2 Summary of the evolution of the figure of merit *ZT* at room temperature ($T = 295$ K), with doping monitored through the conductivity increase with photo- and impurity doping. The color map matches the one in Figure 1. The dashed line marks $ZT = 3$, a value attractive for applications. This requires a resistivity in the $1 \cdot 10^{-3}$ to $3 \cdot 10^{-5}$ Ω cm range (yellow colored zone), accessible by appropriate doping.